\documentclass[
 preprint,
 footinbib,
 showkeys,
 amsmath,
 amssymb,
 aps,
 prb,
 floatfix,
 superscriptaddress,longbibliography
 ]{revtex4-2}

\usepackage{graphicx}
\usepackage{multirow}
\usepackage{xcolor}
\usepackage{dcolumn}
\usepackage{bm}
\usepackage{dsfont}
\usepackage{physics}
\usepackage{hyperref}

\newcommand{\bsigma}[0]{\bm{\sigma}}

\newcommand{\bE}[0]{\bm{\mathcal{E}}}

\DeclareMathOperator{\arccoth}{arccoth}

\begin{document}

\author{Benjamin A.~Levitan}
\email{benjamin.levitan@usherbrooke.ca}
\affiliation{Department of Condensed Matter Physics, Weizmann Institute of Science, Rehovot 7610001, Israel}
\affiliation{Institut Quantique and D\'{e}partement de Physique, Universit\'{e} de Sherbrooke, Sherbrooke J1K 2R1 QC, Canada}
\author{Yuval Oreg}
\author{Erez Berg}
\affiliation{Department of Condensed Matter Physics, Weizmann Institute of Science, Rehovot 7610001, Israel}

\title{Anomalous currents and spontaneous vortices in spin-orbit coupled superconductors}
\begin{abstract}
    We propose a mechanism which can generate supercurrents in spin-orbit coupled superconductors with charged magnetic inclusions. The basic idea is that through spin-orbit interaction, the in-plane electric field near the edge of each inclusion appears to the electrons as an effective spin-dependent gauge field; if Cooper pairs can be partially spin polarized, then each pair experiences a nonzero {\it net} transverse pseudo-gauge field. We explore the phenomenology of our mechanism within a Ginzburg-Landau theory, with parameters determined from a microscopic model. Depending on parameters, our mechanism can either enhance or reduce the total magnetization upon superconducting condensation. Given an appropriate distribution of inclusions, we show how our mechanism can generate superconducting vortices without any applied orbital magnetic field. Our mechanism can produce similar qualitative behavior to the ``magnetic memory effect'' observed in 4Hb-TaS$_2$ \cite{persky2022memory}. However, the magnitude of the effect in that material seems larger than our model can naturally explain.
\end{abstract}

\maketitle

\section{Introduction}
The most dramatic manifestation of the superconducting state is its electromagnetic response \cite{meissner1933,london1935electromagnetic,tinkham2004introduction}. As has been known since the early days of the subject, any superconductor exposed to a magnetic flux density will generate dissipationless supercurrents in an effort to screen that flux density. In type-II superconductors, if the applied flux density is sufficiently large, but not so large as to completely destroy the condensate, then flux penetrates through the cores of quantized vortices. These types of superflow, both related to applied flux, are generic features of a charged condensate subject to electromagnetic gauge invariance.

Depending on details, other forms of superflow are possible. In particular, spin-orbit coupling (SOC) intermingles orbital dynamics (such as current flow) with spin dynamics. A variety of superconducting magnetoelectric effects \cite{levitov1985magnetostatics, edelstein1989noncentrosymmetric, edelstein1995magnetoelectric, yip2002scsoc, samokhin2004soc, bauer2012noncentrosymmetric, pershoguba2015currents, malshukov2016nonlocal, malshukov2017spininjection, olde2019vorticesferro, rabinovich2019magnetoelectricferro, jiang2019qav, malshukov2020spontaneousvortices, samokhvalov2021spontaneous, samokhvalov2022generation, lin2023qav, lu2023socferro} and related phenomena \cite{mardonov2015beccollapse,sakaguchi2016vortexbec,cameron2019vortexrotation,reinhardt2024vortexratchet} can result, putting new twists on the usual relationship between magnetism and supercurrent. In this paper, we identify a magnetoelectric contribution to the supercurrent near inhomogeneities in superconductors with SOC, which, to the best of our knowledge, has thus far gone unnoticed.

Our basic idea begins with the observation \cite{hatano2007nonabeliansoc} that spin-orbit coupling of the form $\mathcal{H}_{\text{SO}} = \lambda (\bE \cross \bsigma) \vdot \vb{p}$ resembles the minimal coupling between the electron momentum operator $\vb{p}$ and a non-abelian gauge potential. $\bE (\vb{r})$ is the local electric field at position $\vb{r}$, and $\bsigma$ is the vector of Pauli matrices for spin. Considering a layered quasi-two-dimensional system and focusing on the term proportional to the in-plane electric field $\bE_{\|}$, we point out that the non-abelian structure reduces to a $\sigma^z$ matrix, i.e., $(\bE_{\|} \cross \bsigma) \vdot \vb{p} = \sigma^z (\bE_{\|} \cross \hat{\vb{z}}) \vdot \vb{p}$. This term looks precisely like the coupling between $\vb{p}$ and an ordinary $U(1)$ electromagnetic vector potential, but with opposite effective vector potentials for opposite spin projections. If Cooper pairs can be at least partially spin polarized, then each pair experiences a nonzero net effective vector potential $\tilde{\vb{A}}$. If $\curl \vb{\tilde{A}} \ne 0$, then the condensate will respond to $\curl \vb{\tilde{A}}$ just as it would to a \textit{real} magnetic flux, either by screening \`{a} la Meissner, or by forming vortices. In this paper, we show how this mechanism can play out when Rashba SOC [proportional to $\mathcal{E}^z$] and exchange fields applied by frozen local moments [modeled as an effective spatially-varying Zeeman splitting $\vb{b} (\vb{r}) \vdot \bsigma$] combine to facilitate the required Cooper pair spin polarization. We note that Holst \textit{et al.}~studied related physics near the edge of a two-dimensional superconductor in Ref.~\cite{holst2022edge2dsc}.

\begin{figure}[h!]
    \centering
    \includegraphics[width=0.7\textwidth]{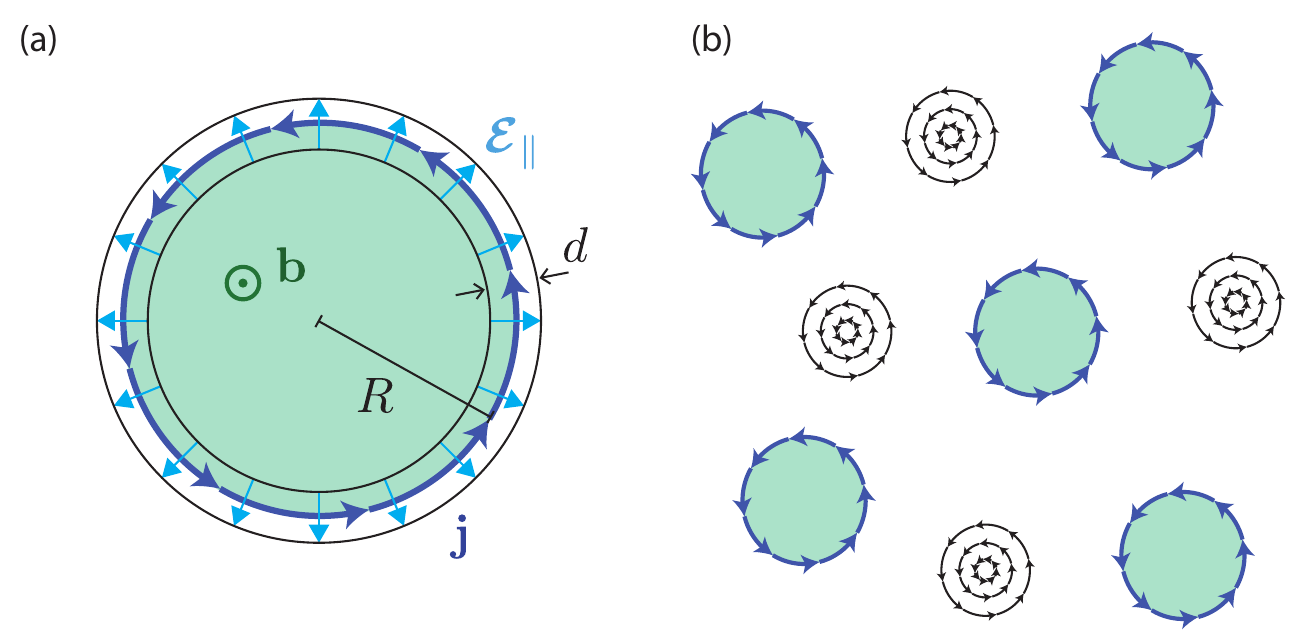}
    \caption{Charged ferromagnetic inclusions in a layered superconductor. (a) A single inclusion. A vertical electric field (not shown) is assumed to be present, providing Rashba coupling; it may be uniform, or random due to charge disorder above and below the plane. The inclusion exerts an out-of-plane exchange field $\vb{b} = b_z \hat{\vb{z}}$ throughout its interior, and a radial electric field $\bE_{\parallel}$ (light blue arrows) in a region of width $d$ around its edge; see Methods section \ref{appendix:electric_field}. A current density $\vb{j}_{\gamma}$ (dark blue) flows in the edge region. (b) Given many inclusions of the type shown in panel (a), vortices can form \textit{in between} the inclusions; see Results section \ref{sec:GL_theory}.}
    \label{fig:geometry}
\end{figure}

In Ref.~\cite{pershoguba2015currents}, Pershoguba \textit{et al.}~studied a related situation of a superconductor subjected to a local exchange field $\vb{b}(\vb{r})$, but with zero in-plane electric field ($\bE_{\|} = 0$). They identified two magnetoelectric supercurrent contributions resulting from Rashba SOC: 
\begin{equation} \label{eq:j_extra}
\vb{j}_{\text{extra}} = \alpha \hat{\vb{z}} \cross \vb{b} + \beta \curl (b_z \hat{\vb{z}}).
\end{equation}
The $\alpha$ term is a well-known magnetoelectric term \cite{levitov1985magnetostatics, edelstein1989noncentrosymmetric, edelstein1995magnetoelectric, yip2002scsoc, bauer2012noncentrosymmetric, samokhin2004soc, pershoguba2015currents, malshukov2016nonlocal}, and the $\beta$ term was pointed out by Pershoguba \textit{et al.}~\cite{pershoguba2015currents}. 
In this work, we identify an additional contribution,
\begin{equation}
    \vb{j}_{\gamma}
    = 
    \gamma \bE_{\|} \cross (b_z \hat{\vb{z}}),
\end{equation}
which arises when the in-plane electric field $\bE_{\|}$ does not vanish, corresponding to local breaking of inversion and rotation symmetries; in homogeneous systems, the $\gamma$ term is forbidden by any point-group symmetry acting nontrivially in the plane (such as inversion or rotation) \cite{smidman2017socsuperconductivity}. Like the $\alpha$ term, and unlike the $\beta$ term, the $\gamma$ term is forbidden in the normal state (see Methods section \ref{appendix:q=0} for details). We note that the $\alpha$ term may be thought of as analogous to the $\gamma$ term, with the electric field oriented out-of-plane and proportional to $\alpha$. \textit{Unlike} the $\alpha$ term, and \textit{like} the $\beta$ term, the $\gamma$ term is even under the mirror $M_z$, so is not forbidden by that symmetry. We analyze the $\gamma$ term in the context of Ginzburg-Landau theory in Results section~\ref{sec:GL_theory}. We propose a scenario where the material contains inclusions which are both charged and magnetic, giving rise to the fields $\vb{b}$ and $\bE$ mentioned above; see Fig.~\ref{fig:geometry}(a). In the superconducting state, supercurrents associated with our $\gamma$ term contribute to the total magnetization. We show that, in the presence of a finite density of such magnetic inclusions and assuming that their magnetizations are aligned, the superconductor may respond by producing vortices located \textit{in between} the inclusions, as shown in Fig.~\ref{fig:geometry}(b). Following the Ginzburg-Landau discussion, we provide a microscopic derivation of the coefficient $\gamma$ in Results section \ref{sec:microscopic}. 

Our study was motivated by anomalous observations of 4Hb-TaS$_2$ \cite{ribak2020chiral,persky2022memory,nayak2023firstorder,almoalem2024littleparks4hb,silber2024twocomponent,almoalem2024charge}, especially those implying a hidden magnetic memory \cite{persky2022memory}. 
The memory manifests as an unexplained increase in total magnetization, in the form of spontaneous vortices, upon cooling into the superconducting state. Crucially, the observed vortex density is far in excess of any magnetic flux, internal or external, present during superconducting condensation. We discuss the observations in question in Results section \ref{sec:comparison_to_4hb}. We find that our mechanism can indeed produce a significant change in total magnetization upon condensation, albeit with a magnitude likely too small to explain the experiments.

\section{Results}

\subsection{Ginzburg-Landau theory}
\label{sec:GL_theory}

We now provide a Ginzburg-Landau description of our effect. We consider a layered 3D superconductor with spin-orbit coupling, containing charged magnetic inclusions which produce an inhomogenous exchange field $\vb{b} (\vb{r})$ and electric field $\bE (\vb{r})$. We assume that the magnetic moments of the inclusions are polarized out-of-plane for simplicity, $\vb{b} = b_z \hat{\vb{z}}$; the $\alpha$ term then vanishes, and we ignore it from now on. Deep in the superconducting phase, up to constant terms (which do not contribute to the supercurrent), the free-energy density is
\begin{equation}    \label{eq:GL_free_energy_density}
    \mathcal{F} [\theta, \vb{A}; \tilde{\vb{A}}]
    =
    \frac{\kappa}{2} \left(
        \grad \theta + 2 e \vb{A} + 2 e \tilde{\vb{A}}
    \right)^2
    +
    \frac{1}{2 \mu_0} \left( \curl \vb{A} \right)^2
\end{equation}
where $-e < 0$ is the electron charge, $\kappa$ is the superfluid stiffness, $\theta$ is the phase of the order parameter, $\vb{A}$ is the electromagnetic vector potential, and
\begin{equation}
    \tilde{\vb{A}}
    =
    \frac{- \gamma}{\kappa (2 e)^2} \bE_{\|} \cross (b_z \hat{\vb{z}})
\end{equation}
is the spin-orbit induced pseudo-gauge field experienced by a Cooper pair \footnote{Note that $\tilde{\vb{A}}$ is an \textit{abelian} pseudo-gauge field. It is a consequence of, but not equal to, the \textit{nonabelian} spin-orbit pseudo-gauge field $\sim \bE \cross \bsigma$.}. $\bE_{\|}$ is the in-plane component of the local electric field $\bE$. At this stage, $\tilde{\vb{A}}$ is motivated by the physical intuition we described in the introduction; we will justify it from a microscopic model below. Note that we treat $\tilde{\vb{A}}$ as externally imposed, while $\vb{A}$ is determined by minimizing the free energy. The supercurrent density corresponding to the free energy in Eq.~\eqref{eq:GL_free_energy_density} is
\begin{equation}
    \vb{j}
    =
    - 2 e \kappa \left(
        \grad \theta + 2e \vb{A} + 2e \tilde{\vb{A}}
    \right).
\end{equation}

We now consider the geometry depicted in Fig.~\ref{fig:geometry}(a). We suppose that a charged magnetic inclusion constitutes a disk of radius $R$ with some vertical thickness. The inclusion produces a net radial electric field $\bE_{\parallel}$ in a region of width $d$ at the edge of the inclusion (we justify this field configuration in Methods section \ref{appendix:electric_field}). The electric field corresponds to a potential difference $V_{\|}$ between the inside and outside of the inclusion, i.e., $\bE_{\parallel} \sim - (V_{\|} / d) \hat{\vb{r}}$ within the edge region. Neglecting variation of $\vb{b}$ across the edge for simplicity, the resulting $\tilde{\vb{A}}$ in the edge region is uniform and purely azimuthal: $\tilde{\vb{A}} = - \gamma b_z V_{\|} \hat{\boldsymbol{\varphi}} / (4 e^2 \kappa d)$. The total $\gamma$-term current flowing around the edge of the inclusion, per unit vertical thickness, is then
\begin{equation}
    I_{\gamma}
    =
    d \vb{j}_{\gamma} \vdot \hat{\boldsymbol{\varphi}}
    =
    \gamma b_z V_{\|}.
\end{equation}
The corresponding magnetic moment per unit vertical thickness is
\begin{equation}    \label{eq:m_gamma}
    M_{\gamma}
    = \pi R^2 I_{\gamma}
    = \pi R^2 \gamma b_z V_{\|}.
\end{equation}
Note that this current can flow in either direction around the inclusion, depending on the sign of $V_{\|}$ (i.e.,~the charge of the inclusion), and that of $\gamma$; $M_{\gamma}$ can thus either add to or cancel against the spin magnetic moment, which we consider next.

We compare $M_\gamma$ to the paramagnetic response of the electronic spins in the magnetic inclusion (present in the normal state as well),
\begin{equation}
\label{eq:m_normal}
    M_{\text{n}}
    =
    \pi R^2 \chi b_z
\end{equation}
where $\chi$ is the spin susceptibility. To estimate the scale of $M_{\text{n}}$, we use the Pauli value of $\chi$, yielding
\begin{equation}
\label{eq:m_normal_pauli}
    M_{\text{n}}
    =
    3 \pi^2 R^2 m \mu_{\text{B}} b_z
\end{equation}
where $m$ is the electron mass and $\mu_{\text{B}}$ is the Bohr magneton. There is also the current due to the $\beta$ term in Eq.~\eqref{eq:j_extra}. This term is not required to vanish in the normal state, but it does so in the microscopic model we consider in Results section \ref{sec:microscopic}. Furthermore, within our microscopic model, the current contribution from the $\beta$ term in the superconductor is suppressed relative to the one from our $\gamma$ term by a factor of the characteristic spin-orbit energy over the chemical potential; see Methods section \ref{appendix:ratio_estimate}. Since this factor is likely to be small, we neglect the $\beta$ term in our present estimate.

The extra contribution to the total magnetic moment in the superconducting state relative to that present already in the normal state is therefore
\begin{equation}    \label{eq:unknown_ratio}
    \frac{M_{\gamma}}{M_{\text{n}}}
    =
    \frac{\gamma V_{\|}}{
        3 \pi m \mu_{\text{B}}
    }.
\end{equation}
To estimate this ratio, we need an estimate for the coefficient $\gamma$, which depends on a choice of microscopic model. We derive $\gamma$ for a simple model below in Results section \ref{sec:microscopic}, starting from Eq.~\eqref{eq:microscopic_H}. Before doing so, we address the question of spontaneous vortex formation within the Ginzburg-Landau picture.

Starting from the free-energy density of Eq.~\eqref{eq:GL_free_energy_density}, we may assume that $\tilde{\vb{A}} = \tilde{\vb{A}}_{T}$ is purely transverse; we can always absorb its longitudinal part $\tilde{\vb{A}}_{L}$ with an appropriate gauge transformation $\grad \theta \rightarrow \grad \theta - 2 e \tilde{\vb{A}}_{L}$. With that assumption, we may write $\tilde{\vb{A}}$ as the curl of some field. Assuming $\kappa$ is uniform, we can write 
\begin{equation}    \label{eq:def_h}
    \curl \tilde{\vb{h}} 
    = - 4 e^2 \mu_0 \kappa \tilde{\vb{A}} 
    = \mu_0 \vb{j}_{\gamma}.
\end{equation}
Note that $\tilde{\vb{h}} / \mu_0$ is the magnetization corresponding to the $\gamma$-term current, $\vb{j}_{\gamma} = -4 e^2 \kappa \tilde{\vb{A}}$. We may choose $\tilde{\vb{h}}$ to be nonzero within the magnetic inclusions and zero outside the inclusions. Neglecting boundary contributions and constant terms (see Methods section \ref{appendix:vortices}), the total bulk free energy can be rewritten as
\begin{equation}    \label{eq:total_GL_energy}
    F[\theta, \vb{A}; \tilde{\vb{A}}]
    =
    \int \dd^2 \vb{r} \mathcal{F}
    =
    \int \dd^2 \vb{r}
    \left[
        \frac{\kappa}{2} \left(\grad \theta + 2 e \vb{A} \right)^2
        + \frac{\pi}{e \mu_0} \vb{n}_{v} \vdot \tilde{\vb{h}}
        + \frac{1}{2 \mu_0} (\curl \vb{A} - \tilde{\vb{h}})^2
    \right],
\end{equation}
where
\begin{equation}
    \vb{n}_{v}
    =
    - \frac{1}{2 \pi}
    \curl \grad \theta
\end{equation}
is the density of superconducting vortices. 
The physical configuration is determined by minimizing the free energy over $\vb{A}$ and $\vb{n}_v$. 
Eq.~\eqref{eq:total_GL_energy} has the form of the standard free energy of a type-II superconductor with $\tilde{\vb{h}}$ playing the role of the external magnetic field (neglecting variations of the amplitude of the superconducting order parameter), with an additional term $\propto \vb{n}_{v} \vdot \tilde{\vb{h}}$. 
Here, we have not included the contribution of vortex cores, where the order parameter amplitude is suppressed over scales of the coherence length, giving rise to a positive core energy (and hence a nonzero lower critical field for vortex formation).

It is useful to first consider the situation where $\tilde{\vb{h}}$ is uniform in the bulk. By Eq.~\eqref{eq:def_h}, this corresponds to $\tilde{\vb{A}} \propto \curl \tilde{\vb{h}}$ (and its associated current) taking nonzero values only near the edge; this situation corresponds to $b_z$ being uniform in the bulk, and $\bE_{\|}\ne 0$ near the edge and zero in the bulk. For a state with no vortices ($\vb{n}_{v} = 0$), the free energy is exactly that of a superconductor in a genuine uniform applied field, which is then Meissner screened by the condensate. The bulk free energy of this state is zero. On the other hand, unlike in a usual superconductor subject to a magnetic field, the $\vb{n}_{v} \vdot \tilde{\vb{h}}$ term prevents the formation of vortices in the bulk. To see this, we assume for simplicity that the penetration depth is extremely large, such that $\vb{B} = \curl \vb{A}$ is essentially uniform. In the usual case of a superconductor in an external field, the superconductor can lower its energy by developing a vortex lattice with spatially-averaged vortex density $\overline{\vb{n}_{v}} = - \overline{\curl \grad \theta} / (2 \pi) = 2 e \curl \vb{A} / (2 \pi) = 2 e \tilde{\vb{h}} / (2 \pi)$, where overlines denote spatial averages. However, in the present case, that configuration would yield a positive bulk free-energy density through the term $\pi \overline{\vb{n}_{v}} \vdot \tilde{\vb{h}} / (e \mu_0) = \tilde{\vb{h}}^2 / \mu_0 > 0$.

In an inhomogeneous system, however, we now show that spontaneous vortices can form. Intriguingly, the vortices appear \textit{away from} the magnetic inclusions responsible for their formation. 
We consider the situation depicted in Fig.~\ref{fig:geometry}(b), with many charged magnetic inclusions embedded into a superconductor, whose magnetizations are all aligned along $+\hat{\vb{z}}$. We assume that the inclusions are distributed randomly in 3D. As before, we assume that the penetration depth is sufficiently large (in particular, larger than the distance between the inclusions) so that the magnetic field can be considered to be uniform. Since $\tilde{\vb{h}}$ is only nonzero within the inclusions and the penetration depth is large, the system can lower its energy by developing vortices in between the inclusions. At the location of the vortices $\tilde{\vb{h}} = 0$, so the $\vb{n}_{v} \vdot \tilde{\vb{h}}$ term does not impose any energetic cost on such a vortex \footnote{If the density of inclusions is sufficiently large, the $\vb{n}_{v} \vdot \tilde{\vb{h}}$ term will force the vortex core to bend around the randomly-distributed inclusions along its vertical extent. This bending should not cost too much energy in the regime of interest.}. However, the large penetration depth means that the flux associated with each vortex penetrates through a large region, including some nearby inclusions: this arrangement hence saves energy by reducing the average value of $(\curl \vb{A} - \tilde{\vb{h}})^2$, while keeping the $\vb{n}_{v} \vdot \tilde{\vb{h}}$ term zero. 

Recall now our motivating question: Can this mechanism explain the spontaneous formation of vortices in 4Hb-TaS$_2$? (See also Results section \ref{sec:comparison_to_4hb}). This question reduces to the issue of magnetization enhancement we considered previously. Magnetization $\boldsymbol{\mathfrak{m}}$ acts as an applied field, and the superconducting condensate forms vortices in order to reduce a free energy term $\propto (\curl \vb{A} -  \mu_0 \boldsymbol{\mathfrak{m}})^2$. Each inclusion contributes a magnetic moment $M_{\text{n}} + M_{\gamma}$ estimated above in Eqs.~\eqref{eq:m_gamma},\eqref{eq:m_normal}, and \eqref{eq:m_normal_pauli}, so the average magnetization density would be $(M_{\text{n}} + M_{\gamma}) \rho_{\text{inc}}$, where $\rho_{\text{inc}}$ is the density of inclusions. Hence, the ratio of novel (i.e., $\gamma$ term related) to conventional vortex densities is $M_{\gamma} / M_{\text{n}}$, given by Eq.~\eqref{eq:unknown_ratio}. We now provide a microscopic calculation of the coefficient $\gamma$ appearing in that equation.

\subsection{Microscopic model}
\label{sec:microscopic}

As discussed in the previous sections, the basic ingredients for the mechanism we propose are superconductivity, spin-orbit coupling, and an exchange field. We consider a simple 2D model containing all three:
\begin{subequations}    \label{eq:microscopic_H}
\begin{equation}
    \mathcal{H}
    =
    \mathcal{H}_0 
    + \mathcal{H}_{b}
    + \mathcal{H}_{\text{SO}},
\end{equation}
where
\begin{equation}
    \mathcal{H}_0
    =
    \left( \frac{1}{2m} \vb{p}^2 \tau^z
    - \mu \tau^z
    - \Delta \tau^x \right) \sigma^0
\end{equation}
\begin{equation}
    \mathcal{H}_{b} = -b_z \sigma^z \, \tau^0
\end{equation}
and
\begin{equation}
    \mathcal{H}_{\text{SO}}
    =
    - \lambda e (\bsigma \cross \bE) \vdot  \vb{p} \, \tau^z.
\end{equation}
\end{subequations}
$\vb{p} = \grad / i$ is the momentum operator (we use $\hbar = 1$). We work in the Nambu basis $\Psi^{\dag} = \begin{pmatrix} \psi^{\dag}_{\uparrow} & \psi^{\dag}_{\downarrow} & \psi_{ \downarrow} & -\psi_{\uparrow} \end{pmatrix}$, where $\psi_{\uparrow / \downarrow}^{\dag} (\vb{r}) $ creates an electron with the indicated spin at position $\vb{r}$. $\bE = \bE_{\|} + \mathcal{E}^z \hat{\vb{z}}$ is the local electric field, and $b_z$ is the local vertical exchange field. Note that $e \bE$ has units of [energy] [length$^{-1}$], and $\lambda$ has units of [length$^2$]. Pauli matrices $\sigma^j$ act on spin, and $\tau^j$ act on particle/hole. We assume a gauge where $\Delta > 0$ for convenience. $k_{\text{F}} = \sqrt{2 m \mu}$ is the Fermi momentum at chemical potential $\mu$ absent spin-orbit coupling, with corresponding velocity $v_{\text{F}} = k_{\text{F}} / m$. We will treat $\bE_{\|} (\vb{r})$ perturbatively, so the largest characteristic energy scale of the spin-orbit coupling is $|\lambda e \mathcal{E}^z| k_{\text{F}}$.

Before describing our calculation, let us spend a few words on the electric field $\bE$. The out-of-plane component $\mathcal{E}^z$ is only required in order to mix triplet components into the singlet Cooper pairs produced by $\Delta$, so the effect is independent of the sign of $\mathcal{E}^z$. Then, since the dependence of our effect on $\mathcal{E}^z$ is even, the effect could result from charge disorder corresponding to local fields with significantly nonzero average $\left( \mathcal{E}^z \right)^2$, even if $\mathcal{E}^z$ averages to zero. 

With respect to the in-plane electric field, in our mechanism, we assume that $\bE_{\|}$ is produced by charged inclusions, as we discussed in the Ginzburg-Landau context in Results section \ref{sec:GL_theory}. Since our goal in the present section is to determine the magnitude of the current, but not its precise distribution, we temporarily disregard the inclusion geometry and instead assume uniform fields $b^z$ and $\bE$ for simplicity \footnote{While the spatially-averaged current density is always required to vanish in equilibrium, standard arguments allow us to compute the current response of the superconducting state in the uniform limit. See Methods section \ref{appendix:q=0} for details.}. 

Now for the calculation. We provide here the outline and key results, leaving the detailed steps to Methods sections \ref{appendix:Green_function}, \ref{appendix:current}, and \ref{appendix:current_cancellation}. We begin with the Hamiltonian $\mathcal{H}$ given in Eq.~\eqref{eq:microscopic_H}. Exploiting translation invariance, we then express $\mathcal{H}$ in momentum space, i.e, $\vb{p} = \grad / i \rightarrow \vb{k}$. The free-electron dispersion, relative to the chemical potential, is $\xi_{\vb{k}} = \vb{k}^2 / (2m) - \mu$. We take $\bE = (\mathcal{E}^x, 0, \mathcal{E}^z)$ without loss of generality---this just amounts to choosing coordinates so that the $x$ axis points along $\bE_{\|}$. The $\vb{q} \rightarrow 0$ current-density operator is
\begin{equation} \label{eq:current_op}
    \vb{J}
    =
    \frac{(-e)}{2 L^2} \sum_{\vb{k}} \Psi^{\dag}_{\vb{k}} \left[
        \frac{\vb{k}}{m} \sigma^0 - \lambda e (\bsigma \cross \bE)_{x, y}
    \right] \tau^0 \Psi_{\vb{k}}
    \equiv
    \frac{1}{L^2} \sum_{\vb{k}} \Psi^{\dag}_{\vb{k}} \vb{J}_{\vb{k}} \Psi_{\vb{k}}.
\end{equation}
The subscript $_{x,y}$ indicates that only the in-plane components of the cross product are considered. The capital $\vb{J}$ distinguishes this current operator from the Ginzburg-Landau current $\vb{j}$.

We compute the current perturbatively in $\mathcal{E}^x$ and $b_z$. We start by perturbing around $b_z = 0$, with $\mathcal{E}^x$ temporarily kept non-perturbative, and then expand the resulting expression in $\mathcal{E}^x$. Writing $k = (i \omega_n, \vb{k})$ with $\omega_n = (2n+1) \pi T$ a fermionic Matsubara frequency at temperature $T$, the linear current response to $b_z$ is
\begin{equation}
    \langle \vb{J} \rangle
    =
    \frac{b_z}{2} \frac{T}{L^2} \sum_{k}
    \tr{
        \vb{J}_{\vb{k}} \mathcal{G}_{k} \sigma^z \tau^0 \mathcal{G}_{k}
    }.
\end{equation}
The Green function $\mathcal{G}$ here includes spin-orbit coupling, but not the exchange field $b_z$---see Methods section \ref{appendix:Green_function}. We focus on $T \rightarrow 0$ for simplicity. Assuming that $0 < |\lambda e \mathcal{E}^z | k_{\text{F}} \ll \mu$, where $k_{\text{F}} = \sqrt{2 m \mu}$ is the Fermi momentum at zero spin-orbit coupling, a straightforward calculation (provided in Methods section \ref{appendix:current}) yields
\begin{equation}    \label{eq:microscopic_current}
    \langle \vb{J} \rangle
    =
    \frac{e^2 \lambda m}{4 \pi}
    f \left[
        \left( \frac{\Delta}{\lambda e \mathcal{E}^z k_{\text{F}}}
        \right)^2
    \right]
    \bE_{\|} \cross (b_z \hat{\vb{z}})
\end{equation}
with
\begin{equation}
    f [y^2]
    =
    1 - \frac{y^2}{\sqrt{1 + y^2}} \arccoth ( \sqrt{1 + y^2}).
\end{equation}
In the limit of small $\Delta$ (so that $0 < \Delta \ll |\lambda e \mathcal{E}^z| k_{\text{F}} \ll \mu$), one has $f[(\Delta^2 / (\lambda e \mathcal{E}^z k_{\text{F}})^2] \rightarrow 1$. This yields the simple expression
\begin{equation} \label{eq:limit_current}
    \langle \vb{J} \rangle
    =
    \frac{e^2 \lambda m}{4 \pi} \bE_{\|} \cross (b_z \hat{\vb{z}}).
\end{equation}
In spite of the simplicity of Eq.~\eqref{eq:limit_current}, our result is in fact highly singular. For small but nonzero $\mathcal{E}^z$ and $\Delta$ (in the order of limits given above), the current is independent of both those parameters, but it vanishes if either one is zero. See Methods sections \ref{appendix:current} and \ref{appendix:current_cancellation} for details. We now match our microscopic calculation to the Ginzburg-Landau theory to determine the coefficient $\gamma$ appearing in $\tilde{\vb{A}}$:
\begin{equation}
    \vb{j}_{\gamma}
    =
    -4 e^2 \kappa \tilde{\vb{A}}
    =
    \gamma \bE_{\|} \cross (b_z \hat{\vb{z}})
    =
    \langle \vb{J} \rangle,
\end{equation}
i.e.,
\begin{equation}
    \gamma
    =
    -\frac{e^2 \lambda m}{4 \pi}
    f \left[
        \left( \frac{\Delta}{\lambda e \mathcal{E}^z k_{\text{F}}}
        \right)^2
    \right]
    \xrightarrow{\Delta \ll |\lambda e \mathcal{E}^z| k_{\text{F}}}
    - \frac{e^2 \lambda m}{4 \pi}.
\end{equation}
We conclude that in the given limit,
\begin{equation}    \label{eq:ratio}
    \frac{M_{\gamma}}{M_{\text{n}}}
    =
    -\frac{e^2 \lambda V_{\|}}{12 \pi^2 \mu_{\text{B}}}.
\end{equation}

\begin{figure}
    \centering
    \includegraphics[width=0.5\textwidth]{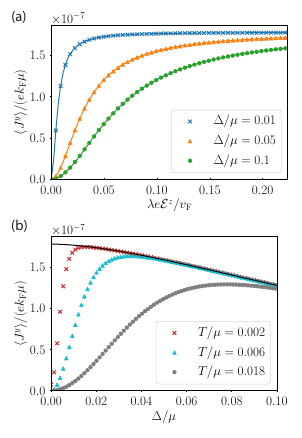}
    \caption{Analytical (curves) and numerical (symbols) results for the current density $\langle J^y \rangle$ given the Hamiltonian of Eq.~\eqref{eq:microscopic_H}, as a function of (a) varying vertical electric field $\mathcal{E}^z$ at fixed superconducting gap $\Delta$, and (b) varying $\Delta$ at fixed $\mathcal{E}^z$. Both panels use $\lambda e \mathcal{E}^x / v_{\text{F}} = -0.01 / \sqrt{20}$ and $b / \mu= 0.001$. All analytical results use the $T \rightarrow 0$ limit. Numerics in panel (a) use $T / \mu = 0.0001$. Panel (b) uses $\lambda e \mathcal{E}^z / v_{\text{F}} = 0.5 / \sqrt{20}$.}
    \label{fig:theory_vs_numerics}
\end{figure}

To extend our results beyond perturbation theory, and to nonzero temperature, we numerically compute the average current $\langle \vb{J} \rangle$ by diagonalizing the $4 \times 4$ Hamiltonian $\mathcal{H}_{\vb{k}}$. Denoting the energy eigenvalues as $\epsilon^{(n)}_{\vb{k}}$ and eigenstates as $\ket{\psi^{(n)}_{\vb{k}}}$, the average current is
\begin{equation}
    \langle \vb{J} \rangle
    =
    \frac{1}{L^2} \sum_{\vb{k}} \sum_{n} \bra{\psi^{(n)}_{\vb{k}}} \vb{J}_{\vb{k}} \ket{\psi^{(n)}_{\vb{k}}}
    n_{\text{F}} \left( \epsilon^{(n)}_{\vb{k}} \right).
\end{equation}
$n_{\text{F}}$ is the Fermi-Dirac distribution at temperature $T$. Fig.~\ref{fig:theory_vs_numerics} displays our analytical and numerical results. As shown in Fig.~\ref{fig:theory_vs_numerics}(b), the singular behavior revealed by our analytical calculation at $T = 0$ is smoothed out at $T > 0$.

\subsection{Comparison to 4H\MakeLowercase{b}-T\MakeLowercase{a}S$_2$}
\label{sec:comparison_to_4hb}
As mentioned in the introduction, our study was motivated by recent experiments on 4Hb-TaS$_2$ \cite{ribak2020chiral,persky2022memory,nayak2023firstorder,almoalem2024littleparks4hb,silber2024twocomponent,almoalem2024charge}. Most relevant to our study is the observation of hidden magnetic memory \cite{persky2022memory}. We begin this section by summarizing that observation. We then argue that while our mechanism can explain some qualitative features of the experiments, the magnitude of the effect predicted by our theory seems too small to account for the experimental observations.

Ref.~\cite{persky2022memory} reported the appearance of spontaneous vortices in the superconducting state of TaS$_2$, despite the lack of any applied magnetic flux during superconducting condensation. 
In particular, vortices were observed if a training field was applied only \textit{above} the superconducting transition temperature $T_{\text{c}}$; the training field was turned off before cooling down through $T_{\text{c}}$.  
The effect persists as long as training is performed below a second temperature $T^* > T_{\text{c}}$. 
The most obvious explanations are an inhomogeneous superconducting transition, or straightforward ferromagnetism present above $T_{\text{c}}$ \cite{greenside1981coexisting,ng1997svp}. 
Both are ruled out by scanning SQUID magnetometry at intermediate temperatures $T_{\text{c}} < T < T^{*}$, which places an upper bound on the magnetization in the normal state: that magnetization is smaller than would be necessary to explain the density of vortices observed in the superconducting state by at least a factor of $10^{3}$.

The question demanding explanation is then: Why does the total magnetization of 4Hb-TaS$_2$ appear to increase drastically when entering the superconducting state? Several answers have been proposed, based on the assumption of: (i) alternating layers of superconductor and either a chiral \cite{lin2024kondospinon} or $\mathds{Z}_2$ \cite{luo2024spontaneous} quantum spin liquid; (ii) unconventional interlayer pairing \cite{liu2024amplification}, or; (iii) a type-II heavy Fermi liquid \cite{konig2024heavyfermi}. No explanation yet has unambiguous experimental support, so the question remains open.

Consider again the scenario depicted in Fig.~\ref{fig:geometry}(b) and discussed in Results section \ref{sec:GL_theory}, where the material hosts a finite density of inclusions which are both charged and magnetic, and which hence exert a local electric field $\bE$ and a local exchange field $\vb{b}$ on the itinerant electrons. In the context of 4Hb-TaS$_2$, we suppose that these inclusions are weakly coupled to one another via the metal into which they are embedded, and that they order (either into a ferromagnet or a glass) at $T^*$. Then, the training field induces some nonzero average exchange field $\vb{b} = b_z \hat{\vb{z}}$, but the associated magnetization is small enough to evade detection in SQUID magnetometry performed above $T_{\text{c}}$. 

In the superconducting state, the spin-orbit coupling induced by $\bE$ then combines with $\vb{b}$ to produce supercurrent through the mechanism we have identified, which (for an appropriate sign of $\gamma$ and the charge of the inclusions) increases the total magnetization below $T_c$. Moreover, if the distance between the inclusions is smaller than the London penetration depth, the superconductor will respond to the magnetic field generated by these supercurrents by creating vortices that are randomly located in between the inclusions. Since the vortices are not pinned to the location of the inclusions, they may appear in different locations on each cooldown (assuming that the inclusions occupy only a small fraction of the sample area, leaving a large area for vortices to explore). Both of these observations are consistent with the experiments: (i) The spontaneous vortices reflect an enhanced magnetization relative to the normal state, and (ii) the locations of the vortices are random and are not pinned to any particular location.

To adequately explain the experiment, however, the magnetization corresponding to the spontaneous vortices generated by our mechanism would need to be several orders of magnitude larger than that corresponding to the magnetization in the normal state. Otherwise, our picture cannot explain why the magnetic order remained invisible when $T_{\text{c}} < T < T^*$. In Methods section \ref{appendix:ratio_estimate}, we estimate the ratio $M_\gamma/M_{\text{n}}$ in Eq.~\eqref{eq:unknown_ratio} and find that it is likely to be at most of order unity. Therefore, the experimentally observed magnetization enhancement in 4Hb-TaS$_2$ seems larger than our mechanism can naturally explain. However, the electronic structure of that material is considerably different from our toy model, for example due to its nonzero Berry curvature, multiple bands crossing the Fermi energy, and strong Ising-type spin-orbit coupling. A full microscopic accounting for these properties lies beyond the scope of this work.

We note that the previously-studied $\alpha$ and $\beta$ terms, $\vb{j}_{\alpha} + \vb{j}_{\beta} = \alpha \hat{\vb{z}} \cross \vb{b} + \beta \curl (b_z \hat{\vb{z}})$, are even more unlikely to explain the experiment in question. In that experiment, the training field is applied along $z$, so one expects $\hat{\vb{z}} \cross \vb{b} = 0$. The $\alpha$ current $\vb{j}_{\alpha}$ then vanishes. The $\beta$ current, $\vb{j}_{\beta}$, does not require in-plane $\vb{b}$, but is unlikely to explain the experiment for two reasons. First, $\beta$ is not required to vanish in the normal state \footnote{In the simple model studied by Pershoguba \textit{et al.}~\cite{pershoguba2015currents}, $\beta$ does happen to vanish in the normal state, but this is not a generic requirement.}. Second, we estimate that under reasonable conditions, the magnetization contribution due to $\beta$ is parametrically smaller than that due to $\gamma$; see Methods section \ref{appendix:ratio_estimate}. 

It is also worth mentioning that Ref.~\cite{jiang2019qav} proposed a mechanism of generating spontaneous vortices bound to magnetic impurities in iron-based superconductors. This mechanism is unlikely to explain the observations in 4Hb-TaS$_2$ since, as mentioned above, the spontaneous vortices in that system do not seem to be pinned to any particular location in the sample. 

\section{Discussion}

We have identified a new mechanism by which charged magnetic inclusions can drive anomalous currents in spin-orbit coupled superconductors. Our results expand the family of known mechanisms by which the interplay of local magnetism and spin-orbit coupling can produce currents in the superconducting state.

Concerning the problem of magnetic memory in 4Hb-TaS$_2$, by assuming that the London penetration depth is much larger than both the size of the inclusions and the distance between them, we argued that our mechanism can indeed drive the spontaneous formation of vortices in between the inclusions, even when the external training field is removed before cooling into the superconducting state. However, within the model we studied, we found that the effect is unlikely to be large enough to explain how the magnetic memory in 4Hb-TaS$_2$ remains hidden in the normal state, given the observed vortex density in the superconducting state. Future work should study whether more realistic electronic structure could yield a larger effect.

Lastly, we point out that the dependence of our mechanism on inclusions suggests a possible experimental signature. Since the spontaneous vortices we have identified avoid the inclusions which produce them, repeated cooldowns should yield different random vortex distributions when the inclusion density is low. Increasing the density of inclusions progressively restricts the low-energy configuration space available to the vortices, ultimately pinning them to fixed locations even on repeated cooldowns, provided that the exchange field configuration does not change.

\section{Methods}
\subsection{The electric field distribution}
\label{appendix:electric_field}
In the Results section \ref{sec:GL_theory}, we assumed a charged inclusion which exerts a radial electric field
near its edge and not elsewhere. In this subsection we justify that assumption. The total screened potential is
\begin{equation}
    V_{\text{tot}, \vb{q}}
    =
    \frac{V_{\text{ext}, \vb{q}}}{
        1 - e^2 U_{\vb{q}} \chi_{\vb{q}}
    },
\end{equation}
where $V_{\text{ext}, \vb{q}} = U_{\vb{q}} \rho_{\text{ext}, \vb{q}}$ is the unscreened potential exerted by the charged inclusion (corresponding to the charge density $\rho_{\text{ext}, \vb{q}}$), $U_{\vb{q}} = 1 / (\epsilon_0 |\vb{q}|)$ is the 2D Fourier transform of the bare Coulomb interaction, and $\chi_{\vb{q}}$ is the static charge response function. In the Thomas-Fermi approximation (valid at small $\vb{q}$, and all the way up to $|\vb{q}| = 2 k_{\text{F}}$ in the 2D Fermi gas), the response function $\chi$ is independent of $\vb{q}$, allowing us to write
\begin{equation}
    V_{\text{tot}, \vb{q}}
    =
    \frac{\rho_{\text{ext}, \vb{q}} / (\epsilon_0 |\vb{q}|)}{
        1 + k_{\text{TF}} / |\vb{q}|
    }
    \approx
    \frac{\rho_{\text{ext}, \vb{q}}}{\epsilon_0 k_{\text{TF}}}.
\end{equation}
The last approximation is valid for momenta much smaller than the Thomas-Fermi inverse screening length, $|\vb{q}| \ll k_{\text{TF}}$. If we assume that the external charge distribution varies slowly compared to the Thomas-Fermi screening length, such that $\rho_{\text{ext}, \vb{q}}$ vanishes when $|\vb{q}|$ approaches $k_{\text{TF}}$, and we have $V_{\text{tot}} \approx \rho_{\text{ext}} / (\epsilon_0 k_{\text{TF}})$ in coordinate space as well. Assuming that the external charge distribution is constant inside the inclusion and decays to zero over a region of width $d \gg 1 / k_{\text{TF}}$, we obtain the in-plane field configuration used in Results section \ref{sec:GL_theory}.

\subsection{The free energy with vortices}
\label{appendix:vortices}
The full Ginzburg-Landau free energy is
\begin{equation}
    F
    =
    \int \dd^2 \vb{r} \mathcal{F}
    =
    \int \dd^2 \vb{r}
    \bigg[
        \frac{\kappa}{2} \left(\grad \theta + 2 e \vb{A} + 2 e \tilde{\vb{A}} \right)^2
        + \frac{1}{2 \mu_0} (\curl \vb{A})^2
    \bigg].
\end{equation}
Using $\tilde{\vb{A}}  = - \curl \tilde{\vb{h}} / (4 e^2 \mu_0 \kappa)$ as in the Results section, we then write
\begin{multline}
    F
    =
    \int \dd^2 \vb{r}
    \bigg[
        \frac{\kappa}{2} \left(\grad \theta + 2 e \vb{A} \right)^2
        -
        \frac{1}{2 e \mu_0} \left(
            \grad \theta + 2 e \vb{A}
        \right) \vdot (\curl \tilde{\vb{h}})
        \\
        +
        \frac{1}{8 e^2 \mu_0^2 \kappa} (\curl \tilde{\vb{h}})^2
        +
        \frac{1}{2 \mu_0} (\curl \vb{A})^2
    \bigg].
\end{multline}
We integrate the second term by parts and 
neglect the resulting surface term (justified since we assume that $\tilde{\vb{h}}$ vanishes outside the material). Dropping the constant (non-dynamical) third term $\sim (\curl \tilde{\vb{h}})^2$, we find
\begin{equation}
    F
    =
    \int \dd^2 \vb{r}
    \left[
        \frac{\kappa}{2} \left(\grad \theta + 2 e \vb{A} \right)^2
        -
        \frac{1}{2 e \mu_0} \tilde{\vb{h}} \vdot \curl \left(
            \grad \theta + 2 e \vb{A}
        \right)
        +
        \frac{1}{2 \mu_0} (\curl \vb{A})^2
    \right].
\end{equation}
Writing the vortex density $\vb{n}_{v} = -\curl \grad \theta / (2 \pi)$, collecting terms, and freely adding a constant term $\sim \tilde{\vb{h}}^2$, we arrive at Eq.~\eqref{eq:total_GL_energy}:
\begin{equation}
    F
    =
    \int \dd^2 \vb{r}
    \left[
        \frac{\kappa}{2} \left(\grad \theta + 2 e \vb{A} \right)^2
        + \frac{\pi}{e \mu_0} \vb{n}_{v} \vdot \tilde{\vb{h}}
        + \frac{1}{2 \mu_0} (\curl \vb{A} - \tilde{\vb{h}})^2
    \right].
\end{equation}

\subsection{Current in the $\vb{q} \rightarrow 0$ limit}
\label{appendix:q=0}
On general grounds, there can never be a nonzero spatially-averaged current in equilibrium, so computing the current under a uniform field configuration as in Results section \ref{sec:microscopic} appears suspicious. However, in the superconducting state, a loophole allows us to estimate the current response by working at $\vb{q} = 0$ in the microscopic model. The argument is the same as for the usual electromagnetic response of a superconductor. First, combine the true vector potential and our effective vector potential into $\vb{A}_{\text{net}} = \vb{A} + \tilde{\vb{A}}$. Within linear response, the current is then given by a response tensor $K_{\vb{q}}$: $\vb{j}_{\vb{q}} = K_{\vb{q}} [\vb{A}_{\text{net}, \vb{q}} + i \vb{q} \theta_{\vb{q}} / (2 e)]$. In a gapped superconducting state, $K_{\vb{q}}$ is continuous and nonzero at $\vb{q} \rightarrow 0$, so we may estimate its value at small but nonzero $\vb{q}$ as approximately equal to $K_{\vb{q} = 0}$. The fact that $K_{\vb{q}=0} \ne 0$ does not violate the requirement of zero net current at equilibrium: for a uniform arrangement of $\vb{A}_{\text{net}}$, the superconducting phase $\theta$ will vary linearly in space such that $2 e \vb{A}_{\text{net}} + \grad \theta = 0$. Our microscopic calculation is then justified: we assume a uniform arrangement of $\bE$ and $\vb{b}$ in the microscopic model, and compute the current neglecting the $\grad \theta$ contribution (which cancels out the $\vb{q} = 0$ current in reality).

The above argument also explains why the $\alpha$ and $\gamma$ terms, corresponding to $\vb{j}_{\alpha} = \alpha \hat{\vb{z}} \cross \vb{b}$ and $\vb{j}_{\gamma} = \gamma \bE_{\|} \cross (b_z \hat{\vb{z}})$, are forbidden in the normal state: without the $\grad \theta$ contribution to the current, both of these terms would allow for nonzero spatially-averaged current in equilibrium. On the other hand, the $\beta$ term is permitted in the normal state: $\vb{j}_{\beta} = \beta \curl (b_z \hat{\vb{z}})$ contributes no current at $\vb{q} = 0$, due to the presence of the curl operator.

\subsection{The Green function at zero exchange field}
\label{appendix:Green_function}
At zero exchange field ($b_z = 0$), one can express the Green function cleanly as
\begin{equation}
    \mathcal{G}_{k}
    =
    \sum_{\alpha = \pm} \Pi_{\vb{k} \alpha} \mathcal{G}_{k \alpha},
\end{equation}
with projectors
\begin{equation}
    \Pi_{\vb{k} \alpha}
    =
    \frac{1}{2} \left(
        \sigma^0 + \alpha \hat{\vb{d}}_{\vb{k}} \vdot \bsigma
    \right) \tau^0
\end{equation}
and Green function components
\begin{equation}
    \mathcal{G}_{k \alpha}
    =
    \sigma^0 \frac{
        i \omega_n \tau^0 + \xi_{\vb{k} \alpha} \tau^z - \Delta \tau^x
    }{(i \omega_n)^2 - E_{\vb{k} \alpha}^2}.
\end{equation}
The spin-orbit pseudovector is $\vb{d}_{\vb{k}} = \lambda e \bE \cross \vb{k}$, the spin-orbit-coupled band energies (relative to $\mu$) are
\begin{equation}
    \xi_{\vb{k} \alpha} = \frac{\vb{k}^2}{2 m} - \mu
    - \alpha |\vb{d}_{\vb{k}}|,
\end{equation}
and the Bogoliubov quasiparticle energies are
\begin{equation}
    E_{\vb{k} \alpha} = \sqrt{\xi_{\vb{k} \alpha}^2 + \Delta^2}.
\end{equation}
$\alpha = \pm$ labels the spin-orbit-coupled bands according to their eigenvalues of $\hat{\vb{d}} \vdot \bsigma$.

\subsection{Computing the current}
\label{appendix:current}
The trace in $\langle \vb{J} \rangle$ factorizes over particle/hole and spin indices using the representation of $\mathcal{G}$ from Methods section \ref{appendix:Green_function}:
\begin{widetext}
\begin{multline}
    \langle \vb{J} \rangle
    =
    -e \frac{b_z}{4} \frac{T}{L^2} \sum_{k}
    \tr{
        \left[
            \frac{\vb{k}}{m} \sigma^0
            -
            \lambda e (\bsigma \cross \bE)_{x, y}
        \right] \tau^0
        \mathcal{G}_{k}
        \sigma^z \tau^0
        \mathcal{G}_{k}
    }
    \\
    =
    -e \frac{b_z}{4} \frac{T}{L^2} \sum_{k} \sum_{\alpha \beta}
    \tr_{\sigma} \left\lbrace
        \left[
            \frac{\vb{k}}{m} \sigma^0
            -
            \lambda e (\bsigma \cross \bE)_{x, y}
        \right]
        \Pi_{\vb{k} \alpha}
        \sigma^z
        \Pi_{\vb{k} \beta}
    \right\rbrace
    \tr_{\tau} \left\lbrace
        \mathcal{G}_{k \alpha} \mathcal{G}_{k \beta}
    \right\rbrace.
\end{multline}
\end{widetext}
The Matsubara summation can be performed using standard methods \cite{altlandsimons}. For the $\alpha = \beta$ terms,
\begin{equation}
    T \sum_{i \omega_n} \tr_{\tau} \left\lbrace
        \mathcal{G}_{k \alpha} \mathcal{G}_{k \alpha}
    \right\rbrace
    =
    n_{\text{F}}' (E_{\vb{k} \alpha})
\end{equation}
Note that at $T \rightarrow 0$, $n_{\text{F}}' (E_{\vb{k} \alpha}) = - \delta(E_{\vb{k} \alpha}) = 0$ as long as $\Delta \ne 0$. In the normal state, the $\delta$-function results in a Fermi surface integral which must be included to obtain the necessary cancellation of the total current; see Methods section \ref{appendix:current_cancellation}. For the $\alpha = \overline{\beta}$ terms,
\begin{equation}
    T \sum_{i \omega_n} \tr_{\tau} \left\lbrace
        \mathcal{G}_{k +} \mathcal{G}_{k -}
    \right\rbrace
    \xrightarrow{T \rightarrow 0}
    \frac{1}{E_{+} + E_{-}}
    \left(
        \frac{\xi_{+} \xi_{-} + \Delta^2}{E_{+} E_{-}} - 1
    \right).
\end{equation}
For the spin sector,
\begin{equation}
    \sum_{\alpha = \pm}
    \tr_{\sigma} \left\lbrace
        \left[
            \frac{\vb{k}}{m} \sigma^0
            -
            \lambda e (\bsigma \cross \bE)_{x, y}
        \right]
        \Pi_{\vb{k} \alpha}
        \sigma^z
        \Pi_{\vb{k} \overline{\alpha}}
    \right\rbrace
    =
    2 e \lambda (\mathcal{E}^z)^2 \mathcal{E}^x \frac{
         k_x k_y \hat{\vb{x}}
         - k_x^2 \hat{\vb{y}}
    }{(\mathcal{E}^z \vb{k})^2 + (\mathcal{E}^x k_y)^2}.
\end{equation}
The $x$ component is odd in $k_{x}$ and $k_y$, so it vanishes in the momentum integral and we may safely ignore it. For the surviving $y$ component, the overall factor of $\mathcal{E}^x$ means that to leading order, we can set $\mathcal{E}^x = 0$ inside the integral, and the only angle dependence in the integrand is through the overall factor of $(k_x / k)^2 = \cos^2 \theta$. We then have
\begin{equation}
    \langle \vb{J} \rangle
    =
    -e^2 \frac{b_z \lambda \mathcal{E}^x}{2}
    \int_{-\pi}^{\pi} \frac{\dd{\theta}}{2 \pi}
    \cos^2 \theta
    \int_{0}^{\infty} \frac{\dd{k} k}{2 \pi}
    \frac{1}{E_{+} + E_{-}}
    \left[
        1 
        - \frac{\xi_{+} \xi_{-} + \Delta^2}{
            E_{+} E_{-}
        }
    \right] \hat{\vb{y}}
\end{equation}
where $\xi_{\pm}$ and $E_{\pm}$ are at $\mathcal{E}^x = 0$. Note that if $\mathcal{E}^z = 0$, then the integrand is identically zero (since $\xi_{+} = \xi_{-}$) and the current vanishes, as mentioned in the Results section. Physically, without the Rashba coupling generated by $\mathcal{E}^z$, the exchange field cannot polarize the singlet Cooper pairs (unless it is strong enough to break them entirely, in which case the current vanishes anyways), and the mechanism we have described in the Introduction and Results sections cannot act.

To evaluate the integral, we linearize around $k_{\text{F}} = \sqrt{2 m \mu}$, writing $k = k_{\text{F}} + k'$ with $k' \ll k_{\text{F}}$. In this case,
\begin{equation}
    \xi_{\pm}
    =
    \frac{1}{2m} (k_{\text{F}} + k')^2
    \mp \lambda e \mathcal{E}^z (k_{\text{F}} + k')
    =
    v_{\text{F}} k' \mp \tilde{v} (k_{\text{F}} + k')
    \approx
    (v_{\text{F}} \mp \tilde{v}) k'
\end{equation}
where the last approximation used $\tilde{v} = |\lambda e \mathcal{E}^z| \ll v_{\text{F}} = k_{\text{F}} / m$.
Then,
\begin{equation}
    \langle \vb{J} \rangle
    \approx
    - \frac{e^2 b_z \lambda \mathcal{E}^x m}{4 \pi}
    f \left[
        \left( \frac{\Delta}{\lambda e \mathcal{E}^z k_{\text{F}}}
        \right)^2
    \right]
    \hat{\vb{y}}
\end{equation}
with
\begin{widetext}
\begin{multline}
    f [y^2]
    =
    \frac{1}{2} \int_{-\infty}^{\infty} \dd{x}
    \frac{
        1 - x^2 - y^2
        - \sqrt{(x - 1)^2 + y^2} \sqrt{(x + 1)^2 + y^2}
    }{
        \sqrt{(x - 1)^2 + y^2} \sqrt{(x + 1)^2 + y^2}
        \left(
            \sqrt{(x - 1)^2 + y^2} + \sqrt{(x + 1)^2 + y^2}
        \right)
    }
    \\
    =
    1 - \frac{y^2}{\sqrt{1 + y^2}} \arccoth ( \sqrt{1 + y^2}).
\end{multline}
\end{widetext}
In the limit $\Delta \ll |\lambda e \mathcal{E}^z| k_{\text{F}}$ we find the result given by Eq.~\eqref{eq:limit_current} in the Results section, $\langle \vb{J} \rangle = -e^2 b_z \lambda \mathcal{E}^x m \hat{\vb{y}} / (4 \pi)$.

\subsection{Current cancellation in the normal state}
\label{appendix:current_cancellation}
Above, we showed that the $T \rightarrow 0$ current also has terms containing the $\delta$ functions $\delta(E_{\vb{k} \alpha})$, which are relevant when $\Delta = 0$ so that $E_{\vb{k} \alpha} = |\xi_{\vb{k} \alpha}|$. The extra piece is then
\begin{widetext}
\begin{multline}
    \langle \vb{J}_{\text{extra}} \rangle
    =
    \frac{e b_z}{4} \int \frac{\dd^2 \vb{k}}{(2 \pi)^2}
    \sum_{\alpha}
    \delta (|\xi_{\vb{k} \alpha}|)
    \tr_{\sigma} \left\lbrace
        \left[
            \frac{\vb{k}}{m} \sigma^0
            -
            \lambda e (\bsigma \cross \bE)_{x, y}
        \right]
        \Pi_{\vb{k} \alpha}
        \sigma^z
        \Pi_{\vb{k} \alpha}
    \right\rbrace
    \\
    =
    \frac{e b_z}{4} \int \frac{\dd^2 \vb{k}}{(2 \pi)^2}
    \sum_{\alpha} \delta (|\xi_{\vb{k} \alpha}|)
    \left[
        \alpha \frac{k_y}{m} \hat{d}^z
        - \lambda e \mathcal{E}^x (\hat{d}^z)^2
        + \lambda e \mathcal{E}^z \hat{d}^x \hat{d}^z
    \right] \hat{\vb{y}}.
\end{multline}
\end{widetext}
As in the $\alpha = \overline{\beta}$ terms, the $x$ component again involves an integrand which is odd in $k_x$ and $k_y$, and hence vanishes. To be consistent with the rest of our calculation, we take the leading-order term in $\mathcal{E}^x$ only, which yields
\begin{equation}
    \langle \vb{J}_{\text{extra}} \rangle
    =
    - \frac{e^2 b_z \lambda \mathcal{E}^x}{4}
    \int \frac{\dd^2 \vb{k}}{(2 \pi)^2}
    \sum_{\alpha} \delta (|\xi_{\vb{k} \alpha}|)
    \left( 
        \frac{k_y}{k}
    \right)^2 \left[
        1
        -
        \alpha \frac{k}{|\lambda e \mathcal{E}^z| m}
    \right] \hat{\vb{y}}.
\end{equation}
The integrand is evaluated at $\mathcal{E}^x = 0$, so the $\delta$ functions pick out the Fermi momenta of the Rashba bands
\begin{equation}
    k_{\text{F}, \alpha}
    =
    \sqrt{2 m \mu + (m \lambda e \mathcal{E}^z)^2}
    +
    \alpha m |\lambda e \mathcal{E}^z|.
\end{equation}
They also give factors of $|\partial \xi_{\vb{k} \alpha} / \partial k|^{-1} = 1 / (v_{\text{F}} - \alpha |\lambda e \mathcal{E}^z|)$, where we assume $|\lambda e \mathcal{E}^z| < v_{\text{F}}$ so that we can drop the overall absolute-value sign in the denominator. In the limit $|\lambda e \mathcal{E}^z| \ll v_{\text{F}}$---the same limit as considered in the superconducting state---we find that the extra contribution in the normal state is
\begin{equation}
    \langle \vb{J}_{\text{extra}} \rangle
    =
    \frac{e^2 b_z \lambda \mathcal{E}^x m}{4 \pi} \hat{\vb{y}},
\end{equation}
which is precisely the negative of the term found before. The total current thus vanishes in the normal state, as it must on general grounds.

\subsection{Estimating the magnetization enhancement} \label{appendix:ratio_estimate}
To estimate the size of the magnetization enhancement we have identified relative to the normal state magnetization, we start from Eq.~\eqref{eq:ratio}. We relate $\lambda$ to the characteristic Rashba energy $\Delta_{\text{R}} = |\lambda e \mathcal{E}^z| k_{\text{F}}$, and express the out-of-plane electric field in terms of its potential drop $\mathcal{E}^z = V_{\perp} / d_{\perp}$. Then, ignoring signs, $\lambda = \Delta_{\text{R}} d_{\perp} / (k_{\text{F}} e V_{\perp})$. Since $\mu_{\text{B}} \sim e/m$, we have
\begin{equation}
    \frac{M_{\gamma}}{M_{\text{n}}}
    \sim
    \frac{V_{\|}}{V_{\perp}} 
    \frac{m \Delta_{\text{R}} d_{\perp}}{k_{\text{F}}}
    \sim
    \frac{V_{\|}}{V_{\perp}} 
    (d_{\perp} k_{\text{F}})
    \frac{\Delta_{\text{R}}}{\mu}.
\end{equation}
We expect $V_{\|} \lesssim V_{\perp}$ (screening is better in the conducting plane than out), $d_{\perp} \sim [1 \text{ to } 10] \times k_{\text{F}}^{-1}$ (fields from impurities are probably not felt very far away) and $\Delta_{\text{R}} < \mu$, so dramatic amplification is unlikely; at most, we expect the extra magnetic moment in the superconducting state to be of similar order to the moment from Pauli paramagnetism.

What about the $\beta$ term? In the limit $\Delta_{\text{R}} \gg m (\lambda e \mathcal{E}^z)^2 \gg \Delta$, $\beta \sim (m \lambda e \mathcal{E}^z)^2 / k_{\text{F}}^2$ \cite{pershoguba2015currents}. In the geometry we consider, the corresponding current $\vb{j} = \beta \grad b_z \cross \hat{\vb{z}}$ yields a magnetic moment
\begin{equation}
    M_{\beta}
    =
    \pi R^2 \beta b_z,
\end{equation}
so
\begin{equation}
    \frac{M_{\gamma}}{M_{\beta}}
    \sim
    \frac{\gamma V_{\|}}{\beta}
    \sim
    \frac{V_{\|}}{V_{\perp}}
    (d_{\perp} k_{\text{F}})
    \frac{\mu}{\Delta_{\text{R}}}.
\end{equation}
Note the factor of $\mu / \Delta_{\text{R}}$, which is likely quite large. In the presence of an in-plane electric field (i.e., around impurities which are both charged and magnetic), our $\gamma$ term then likely plays a larger role than the $\beta$ term identified previously. The $\beta$ term, on the other hand, does not require an in-plane electric field; neutral magnetic impurities suffice. It is interesting to note that for our model, the $\beta$ term vanishes in the normal state, even though this is not required on general grounds. For example, if a term $\propto |\vb{k}|^4$ is added to the bare dispersion relation, then the $\beta$ term survives in the normal state.

\section{Data availability}
The numerical data reported in this work are available upon reasonable request.

\section{Acknowledgements}
The authors acknowledge useful and stimulating discussions with Ehud Altman, Valerio Peri, Gil Refael, and Achim Rosch. B.A.L. was hosted on an extended visit at the Institute for Theoretical Physics at the University of Cologne during most of the preparation of this manuscript, and acknowledges their generous support. B.A.L. also acknowledges the financial support of the Zuckerman STEM Leadership Program. E.B. Acknowledges support from the European Research Council (ERC) under grant HQMAT (Grant Agreement No. 817799) and the Simons Foundation Collaboration on New Frontiers in Superconductivity (Grant SFI-MPS-NFS-00006741-03). 
This work was supported by CRC 183 of the Deutsche Forschungsgemeinschaft (Project C02), and by ISF grants No.~1914/24 and No.~2478/24.

\section{Author contributions}
B.A.L., Y.O., and E.B.~contributed extensively to all aspects of this work.

\section{Competing interests}
The authors declare no competing interests.

\bibliography{4hb}

\end{document}